\documentclass[pss,fleqn]{w-art}
\usepackage{times}
\usepackage{w-thm}
\theoremstyle{plain}

\theoremstyle{definition}

\usepackage[]{graphicx}
\chardef\bslash=`\\ 

\begin{document}

\DOIsuffix{theDOIsuffix}
\Volume{XX}
\Issue{1}
\Month{03} 
\Year{2004}
\pagespan{1}{}
\Receiveddate{15 November 2003}
\Reviseddate{30 November 2003}
\Accepteddate{2 December 2003}
\Dateposted{3 December 2003}
\keywords{Green function, incremental scheme, Landauer theory, molecular junctions.}
\subjclass[pacs]{31.25.-v, 31.25.Jf, 73.40.Ns}



\title{From A Local Green Function to Molecular Charge Transport}


\author[M. Albrecht]{M. Albrecht\inst{1}}\address[\inst{1}]{Theoretical Chemistry FB08, University of Siegen, 57068 Siegen/Germany}
\author[A. Schnurpfeil]{A. Schnurpfeil\inst{1}}
\author[G. Cuniberti]{G. Cuniberti\inst{2}}\address[\inst{2}]{Institute for
  Theoretical Physics, University of Regensburg, 93040 Regensburg/Germany}
%

\begin{abstract} 
A local--orbital based {\it ab initio} approach to obtain the Green
function for large heterogeneous systems is developed. First a Green function formalism is introduced based on 
exact diagonalization. Then the self energy is constructed from an
incremental scheme, rendering the procedure feasible, while at the
same time physical insight into different local correlation
contributions is obtained. Subsequently the Green function is used
in the frame of the Landauer theory and the wide band approximation
to calculate the electronic transmission coefficient across molecular junctions. The theory
is applied to meta-- and para--ditholbenzene linked to gold electrodes
and various
correlation contributions are analyzed.  
\end{abstract}

\maketitle

\section{Introduction}
{\it
}
As of very recently, the trend in miniaturization of electronic
devices indicates a possibility to ultimately arrive at electronically
active elements being constituted by just a single
molecule~\cite{nature,single1}. Specifically it has been demonstrated in break--junction
experiments, that a single organic molecule might be positioned between two
electrodes so as to yield what has become known as a molecular
junction~\cite{goldTip,single2}. Under certain circumstances such a
system allows for
electron transfer from the one electrode to the other. Based on these
experiments two far--reaching conclusions have been ventured, namely
that molecular junctions might lead the way to ultimate one--electron
switches, and secondly, that such a switch behaviour can be triggered
by various rather different physical and chemical circumstances.

However, at present experiments are still notoriously difficult and
hard to interpret, thus shifting weight to theoretical considerations
which are expected to both create a fundamental understanding of the
microscopic processes involved and provide a guidance for improved
engineering.

Theoretical descriptions of the problem try to
illuminate partial aspects like the role of the molecular electronic
structure~\cite{emberly,pantelides1,pantelides2} or the influence of
various structural conformations~\cite{samanta,avouris,haenggi}. Some progress has been made recently on certain applications. It
is the set of methods based on the local density approximation (LDA)
to density functional theory (DFT) as a starting point, which provide numerically
affordable applications to the molecular junction problem~\cite{
emberly,pantelides1,pantelides2,samanta,avouris,haenggi,rakshit_shit}.
LDA--based schemes were pursued by
Rakshit {\it et al}.\cite{rakshit_shit}, Pantelides {\it et
  al}.~\cite{pantelides1,pantelides2} and others~\cite{emberly,samanta,avouris}
for carbon wires and benzene--ring-systems.
Further approximations are introduced by Guti\'errez {\it et\ al.} in an
application to an all--carbon--system with nanotubes as electrodes~\cite{gutierrez}. Along those lines Fagas {\it et al}. analyzed the
  off--resonant electron transport in oligomers~\cite{fagas}, while
  both discussed the switch--like behaviour of a $C_{\rm
    60}$ ball by means of rotation~\cite{gutierrez}. Cuniberti {\it et
  al.} focused on the role of the contacts at the junction~\cite{giovanni_contacts,giovanni_contacts2}. An
  application predicting the actual current-voltage behaviour of two
  aromatic molecules was demonstrated by Heurich {\it et
    al}.~\cite{heurich}. A more advanced scheme developed by Xue and Ratner sticks with these
approximations, but develops a non--equilibrium formalism~\cite{xue_nonequilibrium}. An earlier attempt was presented by Wang
     {\it et
al.}~\cite{wang-nonequilibrium}. 
Further approximations are commonly built
on top of LDA, like the tight binding approach (TB) or parameterized
minimal basis sets~\cite{gutierrez}.
Another set of approaches renounces completely attempts of {\it ab
  initio} calculations and resorts to empirical models~\cite{model,hueckel,scattering,model-compendium}.
Hettler {\it et\ al.} completely abandoned  {\it ab
  initio} approaches and present a semi--quantitative model in
calculations on tunneling transport through benzene~\cite{model}. In
quite the same spirit, several other groups try to tackle the problem
without quantum chemical {\it ab initio} methods, {\it e.\ g.} Ghosh and Datta,
who rely on minimal basis set semi--empirical H\"uckel approaches~\cite{hueckel}. An elastic scattering model is introduced by Todorov
     {\it et
  al.}~\cite{scattering}. A broad compilation of models was written by Paulsson
{\it et al}.~\cite{model-compendium}.
In several applications the Kohn--Sham
eigenenergies are interpreted as excitation energies which is to a
large extent feasible. Principle improvements are available, such as
time dependent DFT~\cite{dft-T} and the optimized effective potential
ansatz~\cite{OEP}. However, it is hard to device systematically
improvable schemes in the frame of DFT.

Wave function based methods, on the other hand, are straightforwardly
applicable to both ground state and excited state calculations
alike and are amenable to systematic improvement on the
numerical accuracy by their very construction. Subsequently, quantities like the DOS or the transport
coefficient $T$ can be obtained in a reliable manner. However, the numerical demand increases
tremendously with the system size. 
Recently it was shown that quantum Monte Carlo techniques (QMC),
particularly diffusion Monte Carlo (DMC), can be applied to excited
states in solids~\cite{QMC-review,williamson,mitas,needsC}.
A second class of approaches extend the coupled cluster (CC) scheme to
excited states. The linear response version of CC was used to obtain
 excitonic transition energies in $Be$, $CH^{\rm +}$, $CO$ and
$H_{\rm 2}O$~\cite{helgaker1}. These approaches can be cast into a reformulation which rests on the
equation--of--motion (EOM) formalism.

The general bottle--neck of steep increase of numerical effort with
system size, however, affects all wave function based methods alike.
It is precisely this obstacle which we strive to overcome by a formulation
of electron correlations in local orbitals and a hierarchy of
correlation contributions called the incremental scheme. In earlier
works we performed band structure calculations with
wave function based {\it ab initio} methods~\cite{gap,toulouse,kiryu,LiF,review}. The key enabling such
calculations for infinite systems was an approach based on local orbitals and a
real space formulation of the self energy. 

In the present work this idea is carried over to the case of
molecules, with applications to large, heterogeneous systems in mind. The purpose is
to display the full theory and introduce the idea of the incremental
scheme in order to construct the Green function. We demonstrate that
this particular combination of
an incremental scheme based on local orbitals and the Green function
theory allows for numerical efficiency and generates an analytic tool
with local resolution. Once the theory is
laid out, the use of the Green function and its underlying incremental
scheme is demonstrated in a simplified application to the problem of
charge transport across a molecular junction. The junction is thought
of as two gold electrodes which are linked via thiol bridges to
meta-- and para--dithiolbenzene. In Sec.\ \ref{sec:theory} the theory
is explained. The numerical results are discussed in Sec.\
\ref{sec:results} and Sec.\ \ref{sec:conclusions} contains our
conclusions.

\section{Theory} 
\label{sec:theory}
\begin{figure}[h]
\centerline{\includegraphics[width=.70\linewidth]{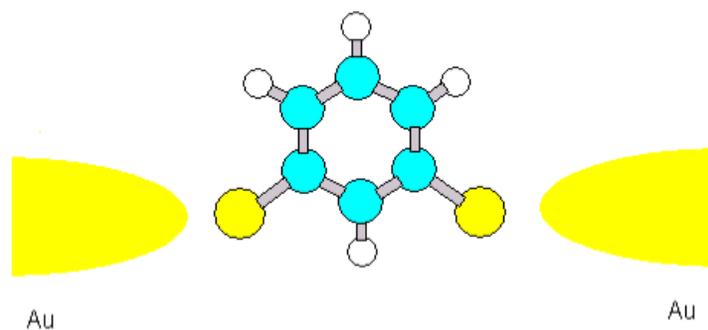}}
\caption{\label{fig:junction}\it Sketch of meta--dithiolbenzene
  illustrating the principle arrangement of so-called molecular junctions.}
\end{figure}

The Landauer theory used to calculate the transmission coefficient
across a molecular junction to be discussed below rests on the calculation of the one--particle Green
function for the region between the junctions. 
Fig.\ \ref{fig:junction} shows a sketch of what we refer to as a
molecular junction. It shows an organic molecule between two
electrodes, which allow to contact the molecule and measure its
current--voltage characteristic. A molecule linked via thiol-bridges
to gold electrodes sets a canonical example for such a system.

Recently we implemented
a wave function based {\it ab initio} procedure to set up the Green
function for band structure calculations in solids and polymers~\cite{kiryu,LiF,review}. Physical insight and numerical efficiency
were derived from an incremental scheme in real space~\cite{gap,toulouse,juergen1,juergen2}.
Here we apply similar ideas to the problem of
molecular junctions. The calculation of the Green function is
described in Sec.\ \ref{subsec:greens}. The combination of the Green
function method with the so called incremental scheme is derived in
Sec.\ \ref{subsec:increment}, while Sec.\ \ref{subsec:landauer}
depicts the Landauer theory. 

\subsection{The Green function} 
\label{subsec:greens}

The starting point of our approach are localized Hartree--Fock (HF)
orbitals for the molecular region between the electrodes, which is in
the simplest case the bare molecule (cf. below). In terms of such
orbitals a model space P and excitation space Q are distinguished for
the example of virtual states (the case of occupied states being
completely analogous) as follows: The model space P describing
the HF level comprises of the ($N+1$)--particle HF determinants   
${\big|}{\eta}{\big\rangle}$, while the
correlation space Q contains single and double excitations ${\big|}{\alpha}{\big\rangle}$ on top of ${\big|}{\eta}{\big\rangle}$:
\begin{eqnarray}
\label{eq-PuQ}
{\big|}{\eta}{\big\rangle}=c_{n}^{\dagger}{\big|}\Phi_{\rm
HF}{\big\rangle},& \qquad &
{\big|}{\alpha}{\big\rangle}=c_{r}^{\dagger}c_{a}{\big|}{\eta}{\big\rangle},
\quad c_{r}^{\dagger}c_{s}^{\dagger}c_{a}c_{b}{\big|}{\eta}{\big\rangle} 
\\
\label{eq-PuQprojectors}
P=\sum_{\eta}{\big|}{\eta}{\big\rangle}{\big\langle}{\eta}{\big|},&  \qquad &
Q=\sum_{\alpha}{\big|}{\alpha}{\big\rangle}{\big\langle}{\alpha}{\big|}  .
\end{eqnarray}
In this local description indices provide an orbital index $n$ which
is normally taken to indicate a local HF orbital.
The idea of locality translates into a restriction
of the area the orbital can be chosen from to one or more contiguous
spatial parts of the molecule. It is important
to note that by enlarging the size of the spatial area thus covered, this approximation
can be checked in a systematic way for convergence. A very important
implication of this procedure is that the correlation methods employed
must be size--consistent. This idea is elaborated upon in Sec.\
\ref{subsec:increment}. 

The Green function approach has precisely the merit of being intrinsically
size--consistent, so that a diagonalization allows to go beyond
perturbative results. Pertaining to the above notation the Green
function  $G_{\rm nm}(t)=-i\langle T[c_{\rm
n}(0)c^{\dagger}_{\rm m}(t)]\rangle$, 
where $T$ is the time--ordering operator and the brackets denote the average
over the exact ground--state, can be obtained from Dyson's equation as:
\begin{equation}
\label{eq-Gkw}
G_{\rm nm}(\omega)=
\left[ 
\omega-F
{\rm
}-\Sigma
(\omega)
\right]^{\rm -1}_{\rm nm}.
\end{equation}
Here the self
energy $\Sigma_{\rm kl}(\omega)$ which contains the correlation
effects, has been introduced. $G^{\rm 0}_{\rm nm}(\omega)$ is the HF propagator
$\left[ G^{\rm 0}(\omega) \right]_{\rm nm}^{\rm -1}={\omega-F_{\rm nm}}$.
The correlated energies are given by the poles of the Green function
which are numerically iteratively retrieved as the zeros of the denominator
in Eq.\ (\ref{eq-Gkw}). To construct the self energy the resolvent 
\begin{eqnarray}
{\left[ \omega-H^{\rm
R}+i0^{+}\right]}_{\rm \alpha\alpha'}^{\rm -1}
\end{eqnarray}
is needed. It can be gained from diagonalization of the Hamiltonian
\begin{eqnarray}
\label{eq-hamiltonian}
[H^{\rm R}]_{\rm \alpha\alpha'}&=&{\langle \alpha\big| H-E_{\rm 0}\big|
\alpha'\rangle},
\end{eqnarray}
where the states $\big|
\alpha\rangle,\big|
\alpha'\rangle$ are those of the correlation space Q as in Eq.\
  (\ref{eq-PuQ}). 

Along with a straightforward implementation several perturbative
approximations have been derived and analyzed. Theoretical connections
to the perturbative effective Hamiltonian~\cite{toulouse} were also established.

\subsection{The incremental scheme}
\label{subsec:increment}
The efficiency of the procedure is derived from the application of an
incremental scheme as follows:
A sketch of the idea is given in Fig.\ \ref{fig-incScheme}. 

As subsets of the incremental scheme some arbitrary spatial parts of a
molecule, representing a suitable partitioning, are chosen for this example. 
An incremental description of the transmission coefficient $T$ could start with a
correlation calculation, in which only excitations inside one of the
regions I--VI, {\it e.\ g.} region I, are allowed. This results in
a contribution to the correlation correction to the
self energy, and ultimately to the transmission coefficient T. With excitations on
top of this being restricted to within region I, the so called one--body increment approximation to the transmission
coefficient is obtained from Eq.\ (\ref{eq:t}) to be discussed in
detail in the next section. Here we just assume, some correlation
method to calculate $T$ would be available. Of course, there
are as many one--body increments as regions chosen to map the
system. They are thus indexed with the region they are referring to. In a next step the
calculation is repeated with  
excitations correlating the charge carriers being allowed to a region enlarged by one
additional part II. The difference with respect to the
one--body increment then isolates the effect of additional excitations involving
this additional region II  and constitutes the two--body increment as shown
in Eq.\ (\ref{eq-oneCell}). This procedure can be continued to more and more
regions. In the end the summation Eq.\ (\ref{eq-pSum}) of all increments is the final approximation
to the sought transmission coefficient.  
\begin{eqnarray}
\label{eq-intra}
{\Delta T^{\rm I}}&=&T^{\rm I}\\ 
\nonumber
\\
\label{eq-oneCell}
{\Delta T^{\rm I,II}}&=&T^{\rm I,II}-{\Delta T^{\rm I}}-{\Delta T^{\rm II}}\\ 
\nonumber
\\
\label{eq-twoCell}
{\Delta T^{\rm I,II,IV}}&=&T^{\rm I,II,IV}-{\Delta T^{\rm I,II}}-{\Delta
T^{\rm I,IV}}- \\ 
\nonumber
& & {\Delta T^{\rm II,IV}}-{\Delta T^{\rm I}}-{\Delta T^{\rm II}}-{\Delta T^{\rm IV}}
\end{eqnarray}
\\
\noindent

\begin{eqnarray}
\label{eq-pSum}
\fbox{$ \displaystyle 
\begin{array}{ccc}
T&=&\sum_{\rm A=I}^{\rm IV}{\Delta{T^{\rm A}}} +\\
\\
& & \sum_{A>B=I}^{\rm IV}{\Delta T^{\rm A,B}}+ \\
\\
& & \sum_{A>B>C=I}^{\rm IV}{\Delta T^{\rm A,B,C}}+ \\
\\
 & & \ .\ .\ .
\end{array}
$}
\end{eqnarray}
\\
\noindent
From the experience gained with the incremental scheme in its application to
the self energy, a rapid
decrease of increments both with the distance between the regions involved
and with their number included in the increment can be expected. This means
that only a few increments need to be calculated. It is crucial to emphasize
that the cutoff thus introduced in the summation Eq.\ (\ref{eq-pSum}) is well
controlled, since the decrease of the incremental series can be explicitly
monitored.  

Furthermore physical information can be extracted from the incremental
scheme. In general the relative weight of specific
increments with respect to others helps to identify important correlation
contributions. 

In earlier applications to band structure calculations it was
demonstrated that the correlation correction $\gamma=(LUMO-HOMO)_{\rm HF}-(LUMO-HOMO)_{\rm CORR}$ of the band gap gives a
suitable measure of the correlation effects accounted for, where
$(LUMO-HOMO)_{\rm HF}$ is the HOMO-LUMO gap on the HF level and
$(LUMO-HOMO)_{\rm CORR}$ is the correlated result. This
correction can also be used as a target quantity for the incremental
scheme in very much the same spirit as the transmission coefficient
inserted above.

\subsection{The Landauer Theory}
\label{subsec:landauer}
From the Green function, the transport coefficient $T$ can be
straightforwardly obtained in the frame of the Landauer formalism~\cite{landauer0}. This
theory constitutes an approximation, assuming zero voltage across the
junction, which is frequently adopted and finds its justification in
the zero-current theorem~\cite{scattering}.

In this context $T$ is given by  

\begin{eqnarray}
\label{eq:t}
 T&=&{\rm Tr}\, \{ \Gamma_{\rm L} \, G \,
\Gamma_{\rm R} \, G^\dagger \} \\
\label{eq:gamma}
\Gamma_{\alpha} &=& i [
\Sigma_{\alpha}-\Sigma^\dagger_{\alpha}] \\
\label{eq:sigma}
\Sigma_{\alpha}&=&H_{\rm M\alpha}G_{\rm \alpha\alpha}^{\rm 0}H_{\rm
  \alpha M}\\
\label{eq:index}
\alpha&=&{\rm L},{\rm R},
\end{eqnarray}
where the indices L,R refer to the coupling to the left (L) and
the right (R) electrode and can be obtained from the self energies
of the respective coupling regions as shown in Eq.\ (\ref{eq:gamma}).
The self energies in turn are obtained in a partitioning approach as
depicted in Eq.\ (\ref{eq:sigma}). Here, the index M refers to the
molecule.
Eq.\ (\ref{eq:sigma}) requires in principle the exact knowledge of the isolated
lead surface Green function $G_{\rm \alpha\alpha}^{\rm 0}$ for both
sides of the junction ($
\alpha={\rm L},{\rm R}$). 
In the wide band approximation, which has been adopted throughout this
work, the coupling
self energies provide an overall widening of the molecular energy
levels, in particular at the thiol bridges, due to the interaction
with the energy continuum provided by the metal~\cite{pantelides1}. This approximation introduces a coupling between
molecule and electrodes parameterized by a coupling strength $\delta$
which replaces the evaluation of Eq.\ (\ref{eq:sigma}).

The Green function $G$ represents the entire system and is to be
obtained from a partition approach leading to:
\begin{eqnarray}
\label{eq:Gges}
G_{\rm MM}&=&{\left[ {G_{\rm MM}^{\rm 0}}^{\rm -1}-\Sigma_{\rm L}-\Sigma_{\rm R} \right]}^{\rm -1},
\end{eqnarray}
where the superscript $0$ refers to the Green function of the bare
molecule, obtained from the {\it ab initio} incremental scheme
to be discussed below in Sec.\ \ref{subsec:increment}. (The scheme is displayed for the key quantity
T, but has been shown in earlier applications to also hold for the
self energy $\Sigma$, hence for the Green function itself, cf. Ref.~\cite{kiryu,LiF,review}).

\section{Results and Discussion} 
\label{sec:results}
The theory presented has been applied to both meta-- and
para--ditholbenzene. A Valence-double-zeta (vdz) basis set with polarization functions was
used throughout the calculations, henceforth denoted as set $A$. For the
purpose of test calculations a smaller basis set (vdz without
polarization functions) denoted as set $B$ was used as well. In a
first preparatory step localized HF orbitals were obtained for the
molecules employing the Pipek-Mezey option of the program package
MOLPRO~\cite{molpro2000}. The four--index transformation was
accomplished by means of the HF program package WANNIER~\cite{alok-wann}
and the subsequent correlation calculations were performed by the
program GREENS developed in our laboratory~\cite{kiryu,LiF}. First the
self energy and the Green function were calculated for the bare
molecule to various levels of accuracy according to the incremental
scheme. The Green function was then inserted into the Landauer
formalism in the frame of the wide band approximation to obtain the
transmission coefficient $T$ across a molecular junction with gold
electrodes.

\subsection{meta--dithiolbenzene}
\label{subsec:meta}

\begin{figure}[h]
\centerline{\includegraphics[width=.70\linewidth]{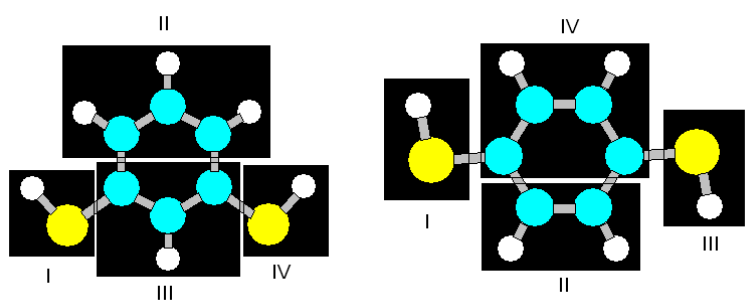}}
\caption{\label{fig-incScheme}\it Sketch of the incremental scheme, exemplifying a possible
  partitioning of the meta--dithiolbenzene (left panel) and para--dithiolbenzene (right panel) comprising four parts denoted as I,
  II, III and IV.}
\end{figure}

The incremental scheme leaves the partitioning of the molecule into
different regions to the user. In the left panel of Fig.\ \ref{fig-incScheme} the
partitioning underlying the following analysis for
meta--dithiolbenzene is depicted. The parts denoted I and IV comprise
the thiol bridges, while parts II and III constitute the carbon
ring. Of course by intuition a partitioning of the carbon ring and its
$\pi$ system would not seem physically reasonable, and indeed this
will be confirmed in the end. At this stage, however, we would like to
demonstrate the usefulness of a local analytic tool in analyzing
precisely such questions. 
%
With this partitioning the incremental scheme is applied both to the
correction $\gamma$ to the HOMO-LUMO gap and the transmission
coefficient $T$. Throughout this work, $T$ is evaluated at the
Fermi--level which was determined by imposing the charge neutrality
condition for the extended system under consideration.\\
\noindent To illustrate the procedure, we first performed calculations
on meta--dithiolbenzene using the basis set B. The results are
summarized in Tab.\ \ref{tab:m_benz_gap_T}. The calculations were done
for a value of 3 eV for the external coupling parameter $\delta$. The
molecule was partitioned into four increments numbered I, II, III and
IV as indicated in the left panel of Fig.\ \ref{fig-incScheme}. The
third row of Tab.\ \ref{tab:m_benz_gap_T} shows the results when we
consider only all one-body increments denoted as 'S'. The correction
for the HOMO-LUMO gap amounts to 2.414 eV if the correlation of the
electrons is calculated by means of MP2 (second order M\o ller-Plesset
theory). This result can be improved when next-neighbored two--body
increments (denoted as 'nn--D') are considered additionally. By
comparing the last two entries in the second column it can be noticed
that including only one--body increments and two--body increments in
the calculation reproduces almost the correlation contribution of the
whole molecule. The same can be noticed when the Epstein-Nesbet theory
to second order (EN2) is applied to calculate the correlation
contribution. The results are depicted in the third column. Here the
values are somewhat increased with respect to the MP2 results, because
beside the Fock contributions of the diagonal elements of $H^{R}$ the
coulomb -and the exchange contributions of the diagonal elements of
$H^{R}$ are included in EN2 while MP2 only considers the Fock
contributions of the diagonal elements of $H^{R}$~\cite{toulouse}. The
calculated values obtained by applying a full diagonalization for all one--body increments is given in the fourth column. 
The results show notedly the local properties of the correlation
contributions, since the one--body increments include the main part of
the correlation energies apart form a little difference which mainly
is included in the two--body increments. The same observation can be
made for the transmission coefficient. The full correlation
contribution (in the sense that all increments are accounted for,
meaning the entire molecule has been correlated) is approximated by
just considering one--body and two--body increments (S and D). So it can be stated that it is unnecessary to include n--body increments with $n>2$. This fact gives rise to an important conclusion:
The numerical effort for the correlation calculation can be lowered
drastically. For this purpose the one--body increments are treated by
full diagonalization because they include the main part of the
correlation while the two--body increments could even be treated by
perturbation theory.

In all of the following calculations, however,
diagonalization is performed for all increments and the basis set A is
employed. Higher ordered increments will again be shown to be of minor
significance. 

For this case we again start by discussing the results for $\gamma$ as shown in Tab.\ \ref{tab:metaBenz2}. Again the row denoted 'S' refers to the result
including all one--body increments while 'S+D' also accounts for all
two--body increments. 

The one--body increments account for most of the correction of the
HOMO-LUMO gap, changing the HF value of 11.619\ eV by an amount of
5.712\ eV. Including all two--body increments brings the HOMO-LUMO gap
down by another 1.711\ eV. These calculations were done without
coupling to the gold electrodes.

\begin{figure}[h]
\centerline{\includegraphics[width=.80\linewidth]{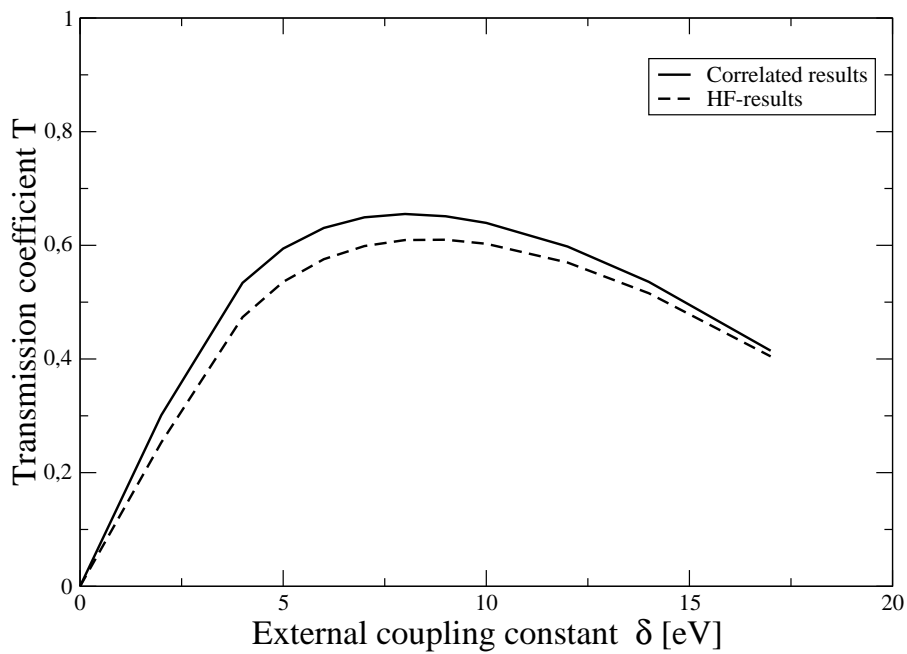}}
\caption{\label{fig:TmetaBenz}\it Transmission coefficient $T$ at the
  Fermi--level for the
  meta--dithiolbenzene in dependence of the contact parameter $\delta$
for the HF-results (dashed) as well as for the correlated results with
one--body-- and two--body increments included (solid).}
\end{figure}

Once the coupling is switched on, the Landauer theory allows to
calculate the transmission coefficient T. In Fig.\ \ref{fig:TmetaBenz}
T is calculated in dependence of the coupling $\delta$. Varying
$\delta$ might be thought of as modeling different geometries at the
junctions continuously, for example the distance between the
electrodes and the molecule. In order to establish the connection to
more realistic descriptions we have run a test calculation to estimate
typical values of the gold--sulfur coupling for a typical
covalent Au--S--distance of 2.38 $\AA$~\cite{AuS} with the program package
MOLPRO~\cite{molpro2000} and found coupling integrals in the range of 4 to 5 eV. The HF results are depicted by dashed
lines while the solid line includes the correlation effects. The
transmission sets on as soon as a finite $\delta$ is chosen and then
increases monotonously with $\delta$ to a peak value at around 8\
eV. With still increasing coupling the transmission then declines
again. The interpretation of this behaviour is as follows: As the
coupling is switched on, the thiol states will hybridize with the
gold atoms, leading to a broadening of the respective molecular states
at the gap, pressing electron density into the former HOMO-LUMO
gap. Conductance then becomes possible. With ever increasing coupling
the broadening of the molecular states then leads to distributing
electronic density into the tales of the spectrum away from the gap
region, thus lowering the transmission again. The overall tendency is
the same for HF and correlated results. The role of the correlation
corrections is to increase the transmission across the junction
quantitatively to some extent. 

Selected typical values of $\delta$ are taken to investigate the
incremental scheme.   
Tab.\ \ref{tab:metaBenz} displays the results for the transmission
coefficient T. The denomination of the rows is as above with 'S'
referring to one--body increments only and 'S+D' including the
two--body corrections as well. Calculations were done for three different
values of the external coupling parameter $\delta$, corresponding to
the increasing branch (4\ eV), the peak region (6\ eV) and the
decreasing branch (10 \ eV) of the graph in Fig.\ \ref{fig:TmetaBenz}. 
(As estimated above a typical covalent situation would be expected to
be situated between 4
and 5 eV).

In all three cases the HF values are corrected to some extent by the
one--body increments. The two--body increments then only give minor
corrections being an order of magnitude smaller. For $\delta=6$\ eV
the HF value of 0.576 is corrected by the one--body increments by an
amount of 0.048 or 8\% to 0.624. The inclusion of two--body increments only
contributes another 0.007. This drop
establishes the rapid convergence of the incremental scheme.

\subsection{para--dithiolbenzene}
\label{subsec:para}
 For the sake of comparison and to elucidate the role of the 
 $\Pi$--system we also did calculations on the para--ditholbenzene,
 which is sketched in the right panel of Fig.\ \ref{fig-incScheme}. Contrary to its
 meta--version it has an inversion symmetry with respect to its center
 of mass. Nonetheless we decided to split the carbon ring again into two
 asymmetric parts, a larger one (region IV, upper part) and a smaller
 one in the lower part, denoted as part II. We would like to use the
 incremental scheme to investigate how the different parts contribute
 to the correlation effects. 
%
%

The results are shown in Tab.\ \ref{tab:paraBenz} for both the
correlation correction $\gamma$ to the HF HOMO--LUMO gap and the
transmission coefficient T. The terms 'S', 'S+D' and 'S+D+T' refer to
inclusion of one--body, two--body and three--body increments. As in
the case of the meta system discussed in the previous section, most
correlation corrections are assessed with the one--body increments,
which correct the HF value of the HOMO-LUMO gap of 11.309\ eV by an
amount of 6.021\ eV. Inclusion of two--body increments leads to an
additional correction of 1.041\ eV, while the three--body increments
together only amount to another -0.003\ eV. This confirms that
converged results are already obtained on the two--body increment
level and justifies the cutoff applied to the incremental series after the
two--body contributions for the calculations on the
meta--dithiolbenzene in the previous section. It is important to note
that the incremental scheme thus allows for a well--monitored and
significant simplification in the numerical effort, since the space to
be diagonalized for two--body increments is in general significantly smaller than
with three regions included. 
As can be seen from the last column in Tab.\ \ref{tab:paraBenz}, these
findings are confirmed for the transmission coefficient $T$ in very much
the same way. As for the case of meta--dithiolbenzene, the convergence
is even faster than for the gap correction. The HF value is changed
from 0.274 by 0.054 or 20\% to 0.328 by the one--body increments
alone. Two--body increments only correct by -0.006 and three--body
increments contribute a meager -0.001.

We now turn to an analysis of the carbon ring contributions. To this
end we check explicitly the combinations of the two--body increments where one part of the
ring is associated with the left thiol--bridge or where both parts are
included and register their influence on the gap correction. The upper and lower part (increments IV and II) are
combined with the left thiol--group (region I in the right panel of Fig.\
\ref{fig-incScheme}) to give the combinations '1+4' and '1+2', respectively,
in Tab.\ \ref{tab:paraBenz}. The improvements on top of the one--body
correlation contributions are virtually negligible, amounting to
0.004\ eV
and zero, respectively. On the other hand, the two--body increment
'2+4', which describes the complete carbon ring, gives a large
contribution of 0.543\ eV. Hence the inclusion of the entire ring
system is essential for a quantitative description. This analysis
shows that the incremental scheme might not only be used as a method
to reduce numerical cost, but could at the same time serve as an
analytic tool with a local resolution.  

\section{Conclusions}
\label{sec:conclusions}

This work introduces an incremental scheme based on local HF orbitals
to construct the self energy and ultimately the Green function with
correlations included in an {\it ab initio} way. The
scheme has been carried over from the case of translationally
invariant systems to heterogeneous systems such as molecular
junctions. The theory combines different ideas into a feasible
procedure and has been put forth in detail. The construction of the Green function and related
quantities was exemplified by calculations on para-- and
meta--dithiolbenzene. The correlation correction of the HOMO-LUMO gap was
discussed in terms of incremental contributions, and a rapid
convergence of the incremental series was found. In the frame of a
simple model, the so called wide band approximation, the molecule was
then linked to gold electrodes, and the Landauer theory was used to
calculate the transmission coefficient across the thus modeled
molecular junction. This simple analysis demonstrates how the
incremental construction of the Green function could be put to use
for the investigation of unconventional systems of high interest. The
rapid convergence of the incremental series when applied to the
transmission was established. Furthermore the usefulness of the
incremental scheme as a local analytic tool became manifest. 

In sum the presented approach allows to efficiently assess correlation
effects in heterogeneous systems such as molecular junctions and
allows to focus on particular parts of the system. We believe the presented
technical analysis allows to start with a series of calculations
comparing the transmission profile of different molecules in the same
environmental context and could thus lead to new insights into the
understanding and engineering of molecular junctions.

\section{Acknowledgements}
\label{sec:acknowledgements}
Valuable discussions with W.\ H.\ E.\ Schwarz (University of Siegen)
and M.\ Dolg (University of K\"oln) are highly appreciated.

\newpage
\begin{table}[h]
\caption{\textit{Correlation correction $\gamma$ of the HF-HOMO-LUMO gap of meta--dithiolbenzene
(Column 2--4. The values are given in eV). $\gamma_{diag}$ labels the
correlation correction obtained by full diagonalization. The last
column contains $T$ at the Fermi-level. Coupling to the electrodes is
accounted for by applying $\delta$ with a value of 3 eV.}}
\label{tab:m_benz_gap_T}
\begin{center}  
\begin{tabular}{|l|l|l|l|l|}
\hline increments: & $\gamma_{\rm MP2}$ & $\gamma_{\rm EN2}$ &
$\gamma_{\rm diag}$ & $T_{\rm MP2}$\\ 
\hline HF (HOMO-LUMO gap: 12.019\hspace{0.1cm}eV)  &  &  & & 0.108 \\ 
\hline all S & 2.414 & 2.789 & 3.900& 0.176  \\ 
\hline all S + nn-D & 2.773 & --- & ---&---  \\
\hline all S + D & 2.919 & 3.436 & ---& 0.179  \\ 
\hline full correlation contribution & 2.921 & 3.439 & 4.636& 0.179 \\ 
\hline
\end{tabular}
  \end{center}   
\end{table}

\newpage
\begin{table}[h]
\protect\caption{\it Incremental scheme for the correlation correction $\gamma$ to
the HOMO-LUMO gap. with
different coupling constants $\delta$ (in eV). Row 'S' gives the correlated
results with all one--body increments included, row 'S+D' also takes
into account all two--body increments.}
\label{tab:metaBenz2}
\begin{center} 
\small{ 
   \begin{tabular}{c|c|c|} 
  \hline     
 \multicolumn{1}{c|}{increment} 
 & \multicolumn{1}{c|}{narrowing of the gap}  &
 \multicolumn{1}{c||}{$\Delta$}    \\ 
  \hline    
  \hline
  HF & HF-gap= 11.619  &  \\
  \hline
  S & 5.712 &${5.712}$  \\
 \hline
  S+D & 7.423 & ${1.711}$  \\
 \hline
  \end{tabular}                      
}
  \end{center} 
\end{table}

\newpage
\begin{table}[h]
\protect\caption{\it Incremental scheme for the transmission
  coefficient $T$ at the Fermi--level with
different coupling constants $\delta$. Row 'S' gives the correlated
results with all one--body increments included, row 'S+D' also takes
into account all two--body increments.}
\label{tab:metaBenz}
\begin{center} 
\small{ 
\begin{tabular}{|l|l|l|l|}
\hline  & $\delta =4.0$ & $\delta =6.0$ & $\delta =10.0$ \\
\hline HF & 0.474 & 0.576 & 0.603 \\
\hline S & 0.526 & 0.624 & 0.635  \\ 
\hline S+D & 0.534 & 0.631 & 0.640 \\  
\hline
\end{tabular}
}
  \end{center} 
\end{table}

\newpage
\begin{table}[h]
\protect\caption{\it Incremental scheme for the transmission
  coefficient $T$ at the Fermi--level and for the correlation correction $\gamma$ to
the HOMO-LUMO gap (in eV) calculated at a 
coupling constants $\delta=2 eV$. Row 'S' gives the correlated
results with all one--body increments included, row 'S+D' also takes
into account all two--body increments while 'S+D+T' states the
  results up to three--body increments. $\Delta$ and $\Delta_{\rm T}$
  monitor the incremental improvement.}
\label{tab:paraBenz}
\begin{center} 
\small{ 
   \begin{tabular}{c|c|c||c|r} 
  \hline     
 \multicolumn{1}{c|}{increment} 
 & \multicolumn{1}{c|}{narrowing of the gap}  &
 \multicolumn{1}{c||}{$\Delta$}  & \multicolumn{1}{c|}{T} &
 \multicolumn{1}{r}{$\Delta_{\rm T}$}  \\ 
  \hline    
  \hline
  HF & gap=11.309  & &  0.274 & \\
  \hline
  all singles & 6.021 &${6.021}$ &  0.328 &0.054 \\
 \hline
  1+2  &6.021 &0.000  &  &  \\
  1+4 &6.024 &0.004 &  &  \\
  2+4 &6.564     &0.543 &  &  \\
\hline
  all doubles & 7.062 & ${1.041}$ & 0.322 &-0.006 \\
 \hline
  all triples & 7.059 & ${-0.003}$ & 0.321 &-0.001 \\
 \hline
  \hline        
  \end{tabular}                      
}
  \end{center} 
\end{table}

\newpage

\newpage

\newpage

\end{document}